\documentclass[reqno,a4paper,11pt,centertags]{amsart}
\usepackage{amsmath,amsthm,amscd,amssymb,latexsym,upref,stmaryrd,cite}
\usepackage{geometry,subfigure}
\usepackage{graphicx}
\usepackage{hyperref}
\newtheorem{theorem}{Theorem}[section]

\newtheorem{proposition}[theorem]{Proposition}

\newtheorem{definition}{Definition}[section]

\title{Synthesis of the Weyl matrix on the square lattice}
\author{Dongjie Wu\textsuperscript{1}}
\author{Chuan-Fu Yang\textsuperscript{2}}
\author{Natalia Pavlovna Bondarenko\textsuperscript{3}}

\begin{document}
	\maketitle
	{
		\footnotetext[1]{Department of Applied Mathematics, School of Mathematics and Statistics, Nanjing University of Science and Technology, Nanjing, 210094, Jiangsu, China, Email: wudongjie@njust.edu.cn}
		\footnotetext[2]{Department of Applied Mathematics, School of Mathematics and Statistics, Nanjing University of Science and Technology, Nanjing, 210094, Jiangsu, China, Email: chuanfuyang@njust.edu.cn}
		\footnotetext[3]{S.M. Nikolskii Mathematical Institute, Peoples' Friendship University of Russia (RUDN University), 6 Miklukho-Maklaya Street, Moscow, 117198, Russian Federation, Email: bondarenkonp@sgu.ru}
	}
	{\noindent\small{\bf Abstract:}
		A method for successive synthesis of the Weyl matrix on the square lattice is proposed. It allows one to compute the Weyl matrix of a large graph by adding new edges and solving elementary systems of linear algebraic equations at each step. Synthesis of the Weyl matrix is useful to further study the inverse problems of the square lattice. Moreover, our approach can be extended to other types of periodic lattices.
	}
	\vspace{1ex}
	
	
	\section{introduction}\label{sec:intro}
\begin{figure}[ht]
		\centering
		\includegraphics[width=4.5cm,height=4cm]{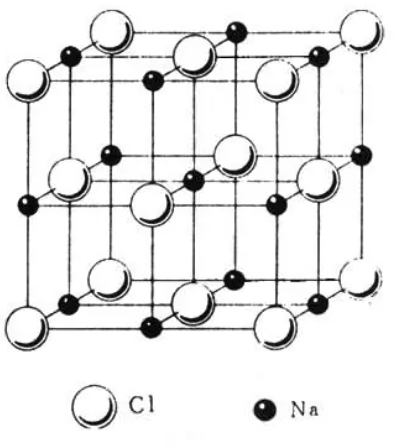}
		\caption{The structure of sodium chloride}
		\label{NaCl}
\end{figure}

	Recently, there have been a lot of studies on quantum graphs, which are essentially spectral problems for the Sturm-Liouville (one-dimensional Schr\"{o}dinger) operators acting on the edges of a graph, while the functions have to satisfy some matching conditions at the vertices. Usually, they are the continuity and the Kirchhoff conditions in internal vertices. In addition, there are some conditions on the boundary. Quantum graphs, or differential equations on metric graphs, have wide applications in nanomaterials, waveguide theory, and chemistry. Interested readers may consult \cite{Ber,Exn,Kur1} for a broad introduction to quantum graphs. The square lattice is a common type of a lattice structure, which has high symmetry. It is widely presented in nature and artificial materials, and has many special physical and chemical properties. The most common example of the square lattice structure is sodium chloride crystals, see figure \ref{NaCl}. In addition to sodium chloride, many metals and compounds also have the square lattice  structure, such as iron, aluminum, copper, and others that are presented in various ceramic materials \cite{Epi,Kes,Loe,Rev,Wu,Yao}. 

\begin{figure}[ht]
		\centering
		\includegraphics[width=7.5cm,height=4cm]{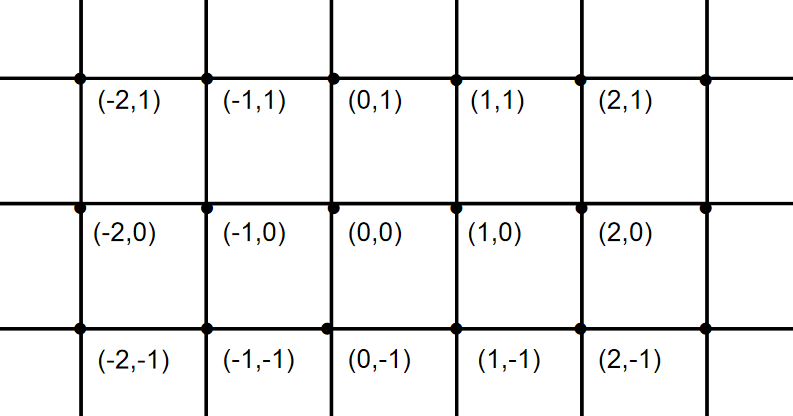}
		\caption{2-dimensional square lattice}
		\label{1}
\end{figure}

By studying the square lattice structure, we can gain a deeper understanding of properties and behavior for materials, and make a contribution to the development of material science and solid-state physics. In the present work, we consider subgraphs of the 2-dimensional square lattice as in figure \ref{1}, which is formed by square having the same edge length such that these square will fit around at each vertex and the pattern at every vertex is isomorphic.

The Weyl matrix of a quantum graph is one of the key mathematical objects, which naturally appears in direct and inverse spectral theory, in the control theory of quantum graphs and in numerous applications, see \cite{And2,And3,Avd1,Avd4,Ber}. By direct problems, we mean investigation of spectral properties (asymptotics, structure, etc.) for operators. Inverse problems consist in the reconstruction of operator coefficients from some spectral information. 
The importance of the Weyl matrix lies in the fact that its transposed matrix is the Dirichlet-to-Neumann (D-N) map of the quantum graph.  For a fixed value of the spectral parameter, a vector of arbitrary Dirichlet-type boundary values of a solution multiplied by the transposed Weyl matrix gives one a vector of the corresponding Neumann-type boundary values if the value of the spectral parameter is not a Dirichlet eigenvalue of the quantum graph (see, e.g., \cite{Avd1}). Moreover, the singularities of the Weyl matrix, considered as a function of the spectral parameter, determine the Dirichlet spectrum of the quantum graph. 

There are many results about inverse problems for quantum graphs exploiting the Weyl matrix or equivalent spectral data, see \cite{Avd1,Bel1,Bel2,Bond20,Bond21, Bro,Fre,Gut,Kur,Yur1,Yur2,Yur3,AK23}. In recent years, spectral theory of differential and difference operators on periodic graphs attracts much attention of scholars. Specifically, direct spectral and scattering problems for Schr\"odinger operators on periodic metric graphs were considered in \cite{Kor1,Kor2,Kuc}. 
Inverse problems on periodic lattices were studied, e.g., in \cite{And1, And2, Iso} for discrete operators and in \cite{And3, And4, WYB} for differential operators. In particular, Wu et al \cite{WYB} have developed a reconstruction procedure for the potentials of the Schr\"odinger operator with a finite support on a square lattice by the D-N map (i.e. the Weyl matrix). However, for numerical experiments with this procedure, the synthesis of the Weyl matrix for the square lattice is crucial.

Direct construction of the Weyl matrix for a sufficiently large quantum graph is quite a challenging problem. It requires solving large systems of equations involving solutions (and their derivatives) of differential equations on all edges of the graph. The main result of the present work is a simple procedure for a progressive synthesis of the Weyl matrix of an arbitrary rectangular subgraph of the square lattice. It refers to the method for a progressive synthesis of the Weyl matrix of an arbitrary quantum tree (i.e. graph without cycles), which has been recently developed by Avdonin et al \cite{Avd4}. 
Thus, staring from a tree-graph, we successively add new edges and compute the Weyl matrices for larger rectangles from the Weyl matrices for the smaller ones.
The proposed synthesis of the Weyl matrix is based on revealed relations between the Weyl solutions of rectangular graphs and their subgraphs. The principal difference of our study from \cite{Avd4} is that adding new edges in our case leads to forming new cycles in the graph. The number of new cycles is sufficiently large at each step, which complicates the procedure. In order to describe our idea, we first consider a simple example of a sufficiently small graph, and then provide a formal description of our algorithm for the general rectangle. Furthermore, since at each step of the procedure we have to solve linear systems, the unique solvability of those systems is discussed.

In the future, our procedure for the synthesis of the Weyl matrix can be applied to the development of numerical algorithms for solving direct and inverse spectral problems and to conducting computational experiments. It is worth mentioning that we have chosen the square lattice for definiteness as the most simple type of a periodic lattice. Undoubtedly, our methods can be specialized to other types of periodic graphs, e.g., to hexagonal lattices.

The paper is organized as follows. In Section \ref{sec:pre}, we give some notations, recall necessary definitions, and express the Weyl solutions in terms of fundamental systems of solutions for the Sturm-Liouville equation on each edge. Section \ref{sec:Syn} gives a simple example of the synthesis of the Weyl matrix. In Section \ref{M-R}, we obtain the synthesis procedure for the rectangular graph of size $m*n$ in the general form. Finally, Section \ref{sec:con} contains concluding remarks.
	
\section{preliminaries} \label{sec:pre}
    Let us define the infinite square lattice $\Gamma=(V,E)$ with the vertex set
	$$
	V := \{ n_1 + \mathrm{i} n_2 \colon n_1, n_2 \in \mathbb Z \} 
	$$
	and the edge set 
	$$
	E := \{ (v,v+1), (v, v+\mathrm{i}) \colon v \in V\}.
	$$
Here, $(v,v+1)$ means that the edge $e=(v,v+1) \in E$ with the end points $v,v+1\in V$. 
	In other words, we consider the vertices as points on the complex plane $\mathbb C$ and suppose that two vertices are joined by an edge if the distance between them equals $1$. For two vertices $w, v \in V$, the notation $w \sim v$ means that there exists an edge $e \in E$ such that $v, w$ are the end points of $e$, and $e \sim v$ means that the edge $e$ is incident to the vertex $v$.
	
	Let $\mathcal E \subseteq E$. Then, we call $f = \{ f_e \}_{e \in \mathcal E}$ \textit{a function} on $\mathcal E$ if each $f_e$ is a function on $e \in \mathcal E$. We will write that $f = \{ f_e \}_{e \in \mathcal E} \in \mathcal A(\mathcal E)$ if $f_e \in \mathcal A[0,1]$ for all $e \in \mathcal E$, where $\mathcal A$ is any functional class, e.g., $\mathcal A = C, L^2$, etc. For $v \in V$, $e \in \mathcal E$, and  $f = \{ f_{e}\}_{e \in \mathcal E}$,  $f_e(v)$ is the value of the function $f$ on $v$, and $\partial_e f(v)=f_e'(v)$ is the value of its derivative in the vertex $v$ along the edge $e$.
	
	Let $G:=\{\mathcal V,\mathcal E \}$ be a finite connected subgraph of the square lattice $\Gamma$. For each vertex $v \in \mathcal V$, put $E_v=\{e \colon e\sim v,e\in \mathcal E\}$. Clearly, $E_v \subseteq \mathcal E$. If $|E_v| = 1$, then $v$ is called a boundary vertex of $\mathcal V$. Denote by $\partial \mathcal V$ the set of the boundary vertices and by $\mbox{int} \, \mathcal V := \mathcal V \setminus \partial \mathcal V$ the set of the internal vertices of $\mathcal V$. 
	The set of the boundary edges $e \in \mathcal E$ such that $e \sim v$, $v \in \partial \mathcal V$ is denoted by $\partial \mathcal E$.
	
	For an internal vertex $v \in \mbox{int} \, \mathcal V$, we say that a function $f = \{ f_e \}_{e \in \mathcal E}$ satisfies \textit{the Kirchhoff conditions} at $v$ if
	
	\smallskip
	
	\textbf{(K-1)} $f$ is continuous at $v$, that is, $f_{e_1}(v) = f_{e_2}(v)$ for any $e_1, e_2 \in E_v$. 
	
	\textbf{(K-2)} $f\in C^1(E_v)$ and $\sum_{e\in E_{v}} \partial_e f(v)=0$.
	
	\smallskip

Let us number the elements of the finite sets $\mathcal V$ and $\mathcal E$
as $\{ \gamma_i \}_{i = 1}^{|\mathcal V|}$ and $\{ e_i \}_{i = 1}^{|\mathcal E|}$, respectively. For convenience, the only edge incident to a boundary vertex $\gamma_i$ is denoted by $e_i$.
Furthermore, we use the notation $f_i(v)=f_{e_i}(v)$ for $i = 1, \dots, |\mathcal E|$ and $v \sim e_i$. Since each edge $e_i$ can be treated as a segment $[0,1]$ and the corresponding element $f_{e_i}$, as a function on $[0,1]$, we also use the notations $f_i(x) = f_{e_i}(x)$ for $x \in [0,1]$.

Let $q = \{ q_i \}_{i = 1}^{|\mathcal E|} \in L^1(\mathcal E)$ be real-valued. Consider the Sturm-Liouville equation on $\mathcal E$:
\begin{equation}\label{S-LEq}
    -\frac{d^2}{dx^2} u_{i}(x)+q_{i}(x)u_{i}(x)=\lambda u_{i}(x),\ on\  e_i \in \mathcal E,
\end{equation}
where $\lambda$ is the spectral parameter.
A function $u$ defined on $\mathcal E$ is said to be a solution of \eqref{S-LEq} on the subgraph $G = (\mathcal V, \mathcal E)$
if besides \eqref{S-LEq} $u$ satisfies (K-1) and (K-2) at each vertex $v\in \mbox{int}\,\mathcal V$.

Now, we give the definition of the Weyl solution and the Weyl matrix as follows.  
	
\begin{definition}
		A solution $w_i$ of \eqref{S-LEq} on the subgraph $G = (\mathcal V, \mathcal E)$ is called the Weyl solution associated with the
		boundary vertex $\gamma_i \in \partial \mathcal V$ if it satisfies the boundary conditions
    \begin{equation}\label{wi}
			w_i(\gamma_i) = 1\   and\   w_i( \gamma_j ) = 0,\  for\  all\  \gamma_j \in \partial \mathcal V,\ j \neq i.
	\end{equation}
\end{definition}

    \begin{definition}
    	The $m\times m$ matrix-function $M(\lambda), \, \lambda \not\in \mathbb R, \, m=|\partial \mathcal V|$, consisting of the elements $M_{ij}(\lambda) =\partial w_i( \gamma_j ), \gamma_j\in \partial \mathcal V$, is called the Weyl matrix of the subgraph $G = \{\mathcal V, \mathcal E\}$. 
    \end{definition}
    
For a fixed value of $\lambda$, the transposed Weyl matrix represents the D-N map of the quantum graph $G$ with the potential $q \in  L^1(\mathcal E)$. Indeed, if $u$ is a solution of \eqref{S-LEq} satisfying
the Dirichlet condition at the boundary vertices $u(\lambda,\partial \mathcal V) = f(\lambda)$, then $\partial u(\lambda,\partial \mathcal V) = M^T(\lambda)f(\lambda)$,
$\lambda\notin \mathbb{R}$. Some properties and applications of Weyl matrix can be found in chapers 17-18 in \cite{Kur1}.

The main result of the present work is a simple procedure which allows one to synthesize the Weyl matrix progressively, by adding edges to smaller graphs and computing the Weyl matrices for the obtained larger graphs from the Weyl matrices for the smaller ones. We call this procedure synthesis of the Weyl matrix.
	
By $\varphi_i(\lambda,x)$ and $S_i(\lambda,x)$, we denote the so-called fundamental solutions of the Sturm-Liouville equation \eqref{S-LEq} on the edge $e_i$, satisfying
the initial conditions:
    $$
    \varphi_i(\lambda,0)=1,\ \varphi'_i(\lambda,0)=0,
    $$ 
	$$
	S_i(\lambda,0)=0,\ S'_i(\lambda,0)=1.
	$$

For a boundary edge $e_i$, it is convenient to identify its boundary vertex $\gamma_i$ with the left endpoint $x = 0$. Then the Weyl
solution $w_i(\lambda, x)$ has the form
	\begin{equation*}
		w_{ii}(\lambda,x)=\varphi_i(\lambda,x)+M_{ii}(\lambda)
S_i(\lambda,x)\ on\  the\  adjacent\  boundary\  edge \  e_i,
	\end{equation*}
    and
    \begin{equation*}
    	w_{ij}(\lambda,x)=M_{ij}(\lambda)S_j(\lambda,x)\ on\  every\  other\  boundary\  edge \  e_j.
    \end{equation*}
    Here, the notation $w_{ij}(\lambda,x)$ means the solution $w_i(\lambda, x)$ on the edge $e_j$.
    
    On internal edges $e_j$, we have
    $$
    w_{ij} (\lambda, x) = a_{ij} (\lambda )\varphi_j (\lambda, x) + b_{ij} (\lambda )S_j (\lambda,x),
    $$
    where the choice of which vertex is identified with zero is arbitrary, and in general the factors
    $a_{ij} (\lambda )$, $b_{ij} (\lambda )$ are unknown.
    
    \section{Simple square graph}\label{sec:Syn}
	In this section, we consider the synthesis of the Weyl matrix for a simple subgraph of the square lattice. This example shows the main idea of our method for construction of the Weyl matrices for arbitrary rectangular graphs. 
		
\begin{figure}[ht]
    	\centering
    	\includegraphics[width=7.5cm,height=6cm]{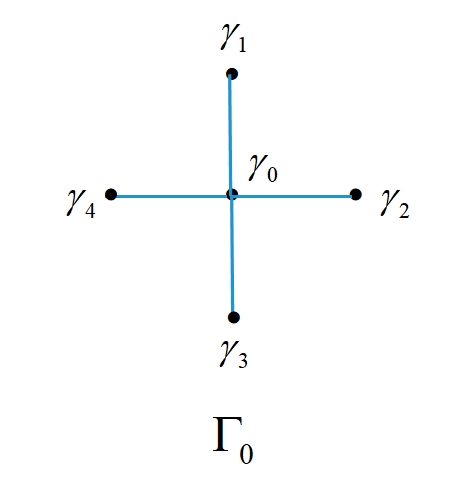}
    	\caption{$0\ast0$ square graph}
    	\label{G0}
    \end{figure}

\begin{figure}[ht]
		\centering
		\includegraphics[width=7.5cm,height=6cm]{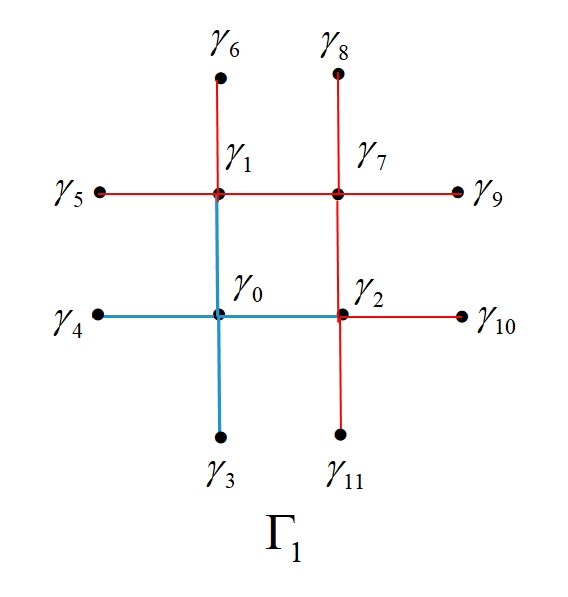}
		\caption{$1\ast 1$ square graph}
		\label{G11}
	\end{figure}

	Consider the simple graph $\Gamma_0$ in figure~\ref{G0}, whose boundary vertices are $\gamma_1,\gamma_2,\gamma_3,\gamma_{4}$. Since $\Gamma_0$ is a tree, its Weyl matrix $\tilde{M}(\lambda)=[\tilde{M}_{ij}(\lambda)]_{i,j=1,\ldots,4}$ can be constructed by the method of \cite{Avd4}.
	Assume $\tilde M(\lambda)$ to be known for some values of $\lambda$. 
	Our task is to find the Weyl matrix $M(\lambda)$ of the $1\ast 1$ square graph $\Gamma_1$ (see figure \ref{G11}). The idea is to add new edges on different boundary vertices step by step and to construct the Weyl solutions $w_i(\lambda, x)$ of $\Gamma_1$ in terms of the Weyl solutions $\tilde{w}_j (\lambda, x)$ of $\Gamma_0$. 
    
\begin{figure}[ht]
		\centering
		\includegraphics[width=12cm,height=6cm]{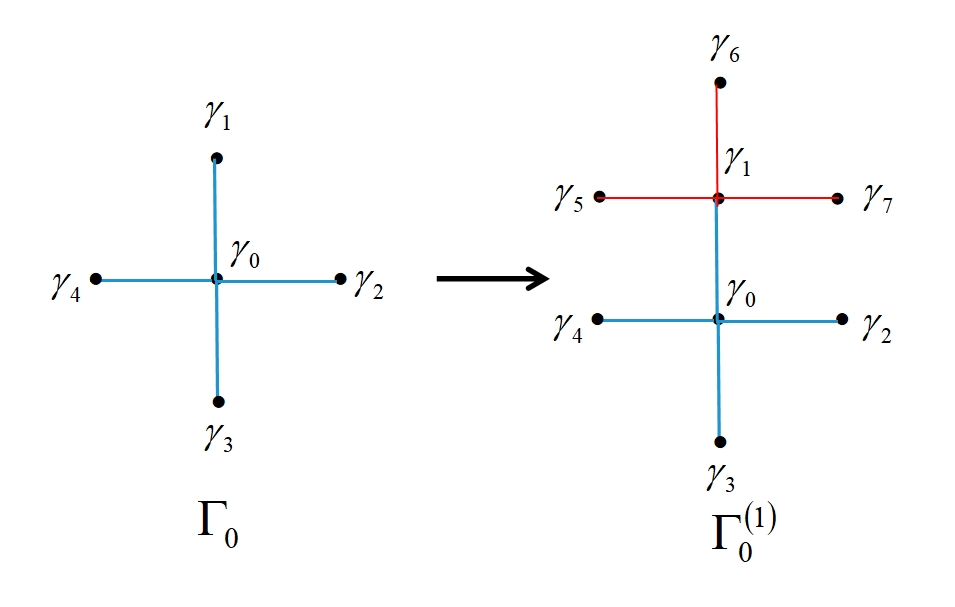}
		\caption{Add edges to $\gamma_1$}
		\label{G0-1}
	\end{figure}

Firstly, we add three edges to the boundary vertex $\gamma_1$ and obtain a new graph $\Gamma_0^{(1)}$, see figure \ref{G0-1}. Since $\Gamma_0$ is a tree, then the Weyl matrix $M^{(1)}(\lambda)$ for it can be easily obtained by the method from \cite{Avd4}.  Namely, let us look for $w^{(1)}_j(\lambda,x),j=5,6,7$ in the form
    \begin{equation}\label{w-w0}
    	 w^{(1)}_j(\lambda,x)=c^{(1)}_j(\lambda)\tilde{w}_1(\lambda,x)\quad  on\ \Gamma_0.
    \end{equation}
That is, on $\Gamma_0^{(1)}$ the Weyl solution $w^{(1)}_j (\lambda, x)$ coincides with $\tilde{w}_1(\lambda, x)$ up to a multiplicative constant $c^{(1)}_j(\lambda)$. In this case, it automatically satisfies the homogeneous Dirichlet condition at $ \gamma_2,\gamma_3,\gamma_4$, and we still need to satisfy the conditions $w^{(1)}_{j}(\lambda,\gamma_j)=1, j=5,6,7$ and $w^{(1)}_{j} (\lambda,\gamma_i)=0$ for $i \neq j$.

By the Kirchhoff conditions on $\gamma_1$, we have
\begin{equation}\label{c0-1}
   \varphi_j(\lambda,\gamma_1)+M^{(1)}_{j,j}(\lambda)S_j(\lambda,\gamma_1)=
   c^{(1)}_j(\lambda),
\end{equation}
\begin{equation}\label{c0-2}
   M^{(1)}_{j,k}(\lambda)S_k(\lambda,\gamma_1)=c^{(1)}_j(\lambda),\quad k\neq j,\ k=5,6,7,
\end{equation}
and
\begin{equation}\label{k-n0}
   \varphi'_j(\lambda,\gamma_1)+\sum_{k=5,6,7}M^{(1)}_{j,k}(\lambda)
   S'_k(\lambda,\gamma_1)=c^{(1)}_j(\lambda)\tilde{M}_{1,1}(\lambda),
\end{equation}
where $\tilde{M}_{1,1}(\lambda)$ is an element of the Weyl matrix $\tilde{M}(\lambda): \tilde{M}_{1,1}(\lambda)= \partial\tilde{w}_1(\gamma_1)$.

Thus, for each $j\in\{5,6,7\}$, from \eqref{c0-1}, \eqref{c0-2}, \eqref{k-n0}, we obtain four equations with the four unknowns $\{ M^{(1)}_{j,k}(\lambda),k=5,6,7;c^{(1)}_j(\lambda) \}$.
Note that in the linear algebraic system \eqref{c0-1}, \eqref{c0-2}, \eqref{k-n0}, the magnitudes $\varphi_k (\lambda,\gamma_1)$, $S_k(\lambda,\gamma_1)$, $\varphi'_k(\lambda,\gamma_1)$, $S'_k(\lambda,\gamma_1)$ are known, since all the potentials $q^{(1)}_k(x),k = 5,6,7$ are known.
Thus, from \eqref{c0-1}, \eqref{c0-2}, \eqref{k-n0}, we find $ M^{(1)}_{j,k}(\lambda),k=5,6,7$, and $c^{(1)}_j(\lambda)$.

\smallskip

\textbf{Unique solvability of the linear algebraic system.}
Let us investigate the unique solvability of the system \eqref{c0-1}, \eqref{c0-2}, \eqref{k-n0}.
First, consider the situation that the potentials $q_k=0$ on all the edges of the graphs $\Gamma_0$ and $\Gamma_0^{(1)}$. Then, the fundamental solutions have the form
\begin{equation}\label{FS}
  \varphi_k(\lambda,x)=\cos \sqrt{\lambda}x,\quad 
S_k(\lambda,x)=\frac{\sin \sqrt{\lambda}x}{\sqrt{\lambda}},\  x\in(0,1),
\end{equation}
on all the edges. For each boundary edge, it is convenient to identify its boundary vertex $\gamma_i$ with the left endpoint $x=0$. The Weyl solution $\tilde{w}_1$ can be represented as 
$$
w_{11}(\lambda,x)=\varphi_1(\lambda,x)+\tilde{M}_{11}(\lambda)
S_1(\lambda,x)=\cos \sqrt{\lambda}x+\tilde{M}_{11}(\lambda)\frac{\sin \sqrt{\lambda}x}{\sqrt{\lambda}},
$$
$$
w_{1k}(\lambda,x)=\tilde{M}_{1k}(\lambda)
S_k(\lambda,x)=\tilde{M}_{1k}(\lambda)\frac{\sin \sqrt{\lambda}x}{\sqrt{\lambda}},\ k=2,3,4.
$$
The Kirchhoff conditions at $\gamma_0$ of the graph $\Gamma_0$ implies
\begin{equation}\label{K1}
   \varphi_1(\lambda,\gamma_0)+\tilde{M}_{11}(\lambda)
   S_1(\lambda,\gamma_0)=\tilde{M}_{1k}(\lambda)
   S_k(\lambda,\gamma_0),\ k=2,3,4,
\end{equation}
and 
\begin{equation}\label{K2}
   \varphi'_1(\lambda,\gamma_0)+\sum_{k=1}^{4}\tilde{M}_{1k}(\lambda)
   S'_k(\lambda,\gamma_0)=0.
\end{equation}
By \eqref{K1},\eqref{K2} and \eqref{FS}, we can compute
$$\tilde{M}_{11}(\lambda)=\frac{\sqrt{\lambda}\sin \sqrt{\lambda}}{4\cos \sqrt{\lambda}}-\frac{3\sqrt{\lambda}\cos \sqrt{\lambda}}{4\sin \sqrt{\lambda}}.
$$
Then, the coefficient matrix $A_0$ of the linear algebraic equations \eqref{c0-1}, \eqref{c0-2}, \eqref{k-n0} has the form
$$
A_0=
\begin{pmatrix}
\frac{\sin \sqrt{\lambda}x}{\sqrt{\lambda}} & 0 & 0 & -1 \\
0 & \frac{\sin \sqrt{\lambda}x}{\sqrt{\lambda}} & 0 & -1 \\
0 & 0 & \frac{\sin \sqrt{\lambda}x}{\sqrt{\lambda}} & -1 \\
\cos \sqrt{\lambda}x & \cos \sqrt{\lambda}x & \cos \sqrt{\lambda}x & -\tilde{M}_{11}(\lambda) 
\end{pmatrix}
$$
After some simple calculations, the determinant of the coefficient matrix $A_0$ can be represented as 
\begin{align*}
\det(A_0)&=3\cos \sqrt{\lambda}\frac{\sin^2 \sqrt{\lambda}}{\lambda}-\tilde{M}_{11}(\lambda)\frac{\sin^3 \sqrt{\lambda}}{\lambda^{\frac{3}{2}}}\\
&=\frac{15}{4}\cos \sqrt{\lambda}\frac{\sin^2 \sqrt{\lambda}}{\lambda}-\frac{\sin^4 \sqrt{\lambda}}{4\lambda\cos \sqrt{\lambda}}.
\end{align*}
We can easily compute the zeros of $\det(A_0)$ as follows
$$
\sqrt{\lambda_k^0}=k\pi,\ k\in\mathbb{Z}\setminus\{0\}, \quad \text{(double zeros)}
$$
and
$$
\sqrt{\lambda_k^{0\pm}}=2k\pi+\arccos\left(\pm\frac{1}{4}\right),\ k\in \mathbb{Z}.
$$
Then, the unique solvability of the linear system \eqref{c0-1}, \eqref{c0-2}, \eqref{k-n0} for some fixed values of $\lambda$ except the zeros of $\det(A_0)$ is proved. 

In the case $q_k \ne 0$, the fundamental solutions have the form
\begin{equation*}\label{FS'}
  \varphi_k(\lambda,x)=\cos \sqrt{\lambda}x+O\left(\frac{e^{|\tau|x}}{\sqrt{\lambda}}\right),\ 
S_k(\lambda,x)=\frac{\sin \sqrt{\lambda}x}{\sqrt{\lambda}}+o\left(\frac{e^{|\tau|x}}{\sqrt{\lambda}}\right),
\end{equation*} 
for $x\in(0,1),\ |\lambda|\rightarrow\infty$ on all edges. Here $\tau=Im \sqrt{\lambda}$, and $o$ and $O$ denote the Landau symbols. Introduce the region
$$
G_\delta=\left\{ \sqrt{\lambda} \colon \Big|\sqrt{\lambda}-\frac{(2k+1)\pi}{2}\Big|\geq \delta,\ k\in \mathbb{Z} \right\},\ \delta > 0.
$$
Then, the determinant of the coefficient matrix $A_1$ can be represented as 
\begin{align}\label{A_1}
&\det(A_1)=3\cos \sqrt{\lambda}\frac{\sin^2 \sqrt{\lambda}}{\lambda}-\tilde{M}_{11}(\lambda)\frac{\sin^3 \sqrt{\lambda}}{\lambda^{\frac{3}{2}}}+o\left(\frac{e^{3|\tau|}}{\lambda}\right)\notag\\
&=\frac{15}{4}\cos \sqrt{\lambda}\frac{\sin^2 \sqrt{\lambda}}{\lambda}-\frac{\sin^4 \sqrt{\lambda}}{4\lambda\cos \sqrt{\lambda}}+o\left(\frac{e^{3|\tau|}}{\lambda}\right),\ \sqrt{\lambda}\in G_\delta, \quad |\lambda| \to \infty.
\end{align}
Consequently, the zeros of $\det(A_1)$ have the asymptotics
$$
\sqrt{\lambda_k^{1,2}} = \sqrt{\lambda_k^0} + o(1), \quad
\sqrt{\lambda_k^{\pm}} = \sqrt{\lambda_k^{0\pm}} + o(1), \quad |k| \to \infty.
$$

Clearly, in the points $\lambda \not\in \{ \lambda_k^{1,2}, \lambda_k^{\pm}\}$, the system \eqref{c0-1}, \eqref{c0-2}, \eqref{k-n0} is uniquely solvable.
Thus, we arrive at the following proposition.

\begin{proposition} \label{prop:solve}
For each $\delta > 0$ there exists a constant $C > 0$ such that 
the linear algebraic system \eqref{c0-1}, \eqref{c0-2}, \eqref{k-n0} is uniquely solvable for each $\lambda$ such that 
$$
|\lambda| \ge C, \quad |\sqrt{\lambda} - k\pi| \ge \delta, \quad
\left| \sqrt{\lambda} - 2k\pi - \arccos\Big( \pm \frac{1}{4}\Big)\right| \ge \delta.
$$
\end{proposition}
	
In particular, Proposition~\ref{prop:solve} justifies our algorithm for the values of $\lambda$ with sufficiently large $|\mbox{Im}{\sqrt \lambda}|$. Analogously, the unique solvability can be studied for the linear algebraic systems in the arguments below.

\smallskip

Proceed with the synthesis of the Weyl matrix $M^{(1)}(\lambda)$. From \eqref{w-w0}, we can obtain 
\begin{equation*}
  M^{(1)}_{j,i}(\lambda)=c^{(1)}_j(\lambda)\tilde{M}_{1,i}(\lambda),\ i=2,3,4,
\end{equation*}
and thus we have already completed the rows $4,5,6$ of $M^{(1)}(\lambda)$.

Now, choose an $i\in \{2,3,4\}$. We look for $w^{(1)}_i(\lambda,x)$ such that on $\Gamma_0$ the following equality is valid:
$$
w^{(1)}_i(\lambda, x) = \tilde{w}_i(\lambda, x) + \alpha^{(1)}_i(\lambda)\tilde{w}_1(\lambda, x),
$$
where $\alpha^{(1)}_i(\lambda)$ is a constant. This is a natural choice, because
$w^{(1)}_i(\lambda,\gamma_i) = 1$ and $w^{(1)}_i(\lambda,\gamma_j) = 0$ for $j = 2,3,4,\  j \neq i$.

By the Kirchhoff conditions at $\gamma_1$, we have 
\begin{equation}\label{c0-3}
  M^{(1)}_{i,j}(\lambda)S_j(\lambda,\gamma_1)=\alpha^{(1)}_i(\lambda),\ j=5,6,7,
\end{equation}
and
\begin{equation}\label{c0-4}
  \sum_{j=5,6,7}M^{(1)}_{i,j}(\lambda)S'_j(\lambda,\gamma_1)=\tilde{M}_{i,1}
  (\lambda) + \alpha^{(1)}_i(\lambda)\tilde{M}_{1,1}(\lambda).
\end{equation}

For every $i\in \{2,3,4\}$, relations \eqref{c0-3}, \eqref{c0-4} give us four equations for the four unknowns $\{ M^{(1)}_{i,j}(\lambda),j=5,6,7, \,  \alpha^{(1)}_i(\lambda) \}$.
The magnitudes $S_j (\lambda,\gamma_1)$, $S'_j(\lambda,\gamma_1)$, $\tilde{M}_{i,1}(\lambda)$, $\tilde{M}_{1,1}(\lambda)$ in \eqref{c0-3}, \eqref{c0-4} are already known. So, we get the elements of the Weyl matrix $M^{(1)}(\lambda)$ in the columns $4,5,6$ of the rows $1,2,3$.

Finally, to obtain $M^{(1)}_{i,j}(\lambda)$ for $i, j = 2,3,4$, that is, the elements of the Weyl matrix $M^{(1)}(\lambda)$ in the columns $1,2,3$ of the rows $1,2,3$.  Observe that
$$
\partial w^{(1)}_i(\lambda,\gamma_j)=\partial \tilde{w}_i(\lambda,\gamma_j)+ \alpha^{(1)}_i(\lambda)\partial \tilde{w}_1(\lambda,\gamma_j),
$$
and so
\begin{equation}\label{M_ij0}
  M^{(1)}_{i,j}(\lambda)=\tilde{M}_{i,j}(\lambda) + \alpha^{(1)}_i(\lambda)\tilde{M}_{1,j}(\lambda)\ for \ i,j=2,3,4.
\end{equation}

Now, we have already completed the construction of the Weyl matrix $M^{(1)}(\lambda)=[M_{ij}^{(1)}(\lambda)]_{i,j=2,\ldots,7}$ for the graph $\Gamma_0^{(1)}$ from Weyl matrix $\tilde{M}(\lambda)$ of the graph $\Gamma_0$ by adding new edges to the boundary vertex $\gamma_1$.

\begin{figure}[ht]
		\centering
		\includegraphics[width=12cm,height=6cm]{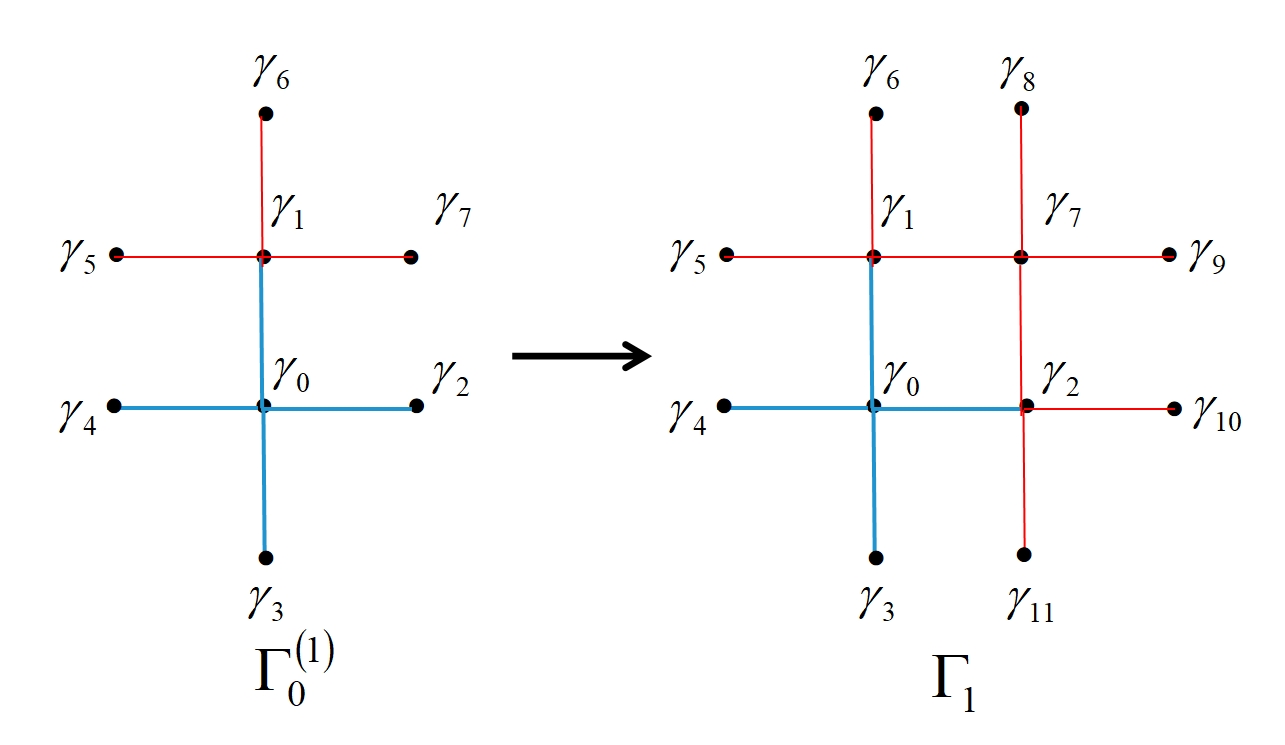}
		\caption{Add edges to $\gamma_2$ and $\gamma_7$}
		\label{G0-2}
\end{figure}

Next, we add new edges to the right boundary vertices of the graph $\Gamma_0^{(1)}$ and obtain the new graph $\Gamma_1$ as in figure~\ref{G0-2}. Our goal is to construct the Weyl matrix $M(\lambda)=M^{(2)}(\lambda)$ of the graph $\Gamma_1$ from the Weyl matrix $M^{(1)}(\lambda)$ of the graph $\Gamma_0^{(1)}$. 

Let $j \in \{10, 11\}$ be fixed. Look for the Weyl solutions $w^{(2)}_j(\lambda,x)$ in the form
\begin{equation}\label{0w-w2}
    w^{(2)}_j(\lambda,x)=
    c^{(2)}_{j,2}(\lambda)w^{(1)}_2(\lambda,x) + c^{(2)}_{j,7}(\lambda)w^{(1)}_7(\lambda,x) \  on \ \Gamma_0^{(1)}.
\end{equation}
That is, on $\Gamma_0^{(1)}$ the Weyl solution $w^{(2)}_j(\lambda,x)$ is a linear combination of $w^{(1)}_2(\lambda, x)$ and $w^{(1)}_7(\lambda, x)$. In this case, it automatically satisfies the homogeneous Dirichlet conditions at $ \{\gamma_3,\cdots,\gamma_{6}\}$, and we still need to satisfy the conditions $w^{(2)}_{j}(\lambda,\gamma_j)=1,\  j=10,11$ and $w^{(2)}_{j} (\lambda,\gamma_i)=0$ for $i \neq j$. The solution on the edge $e_1=(\gamma_2,\gamma_7)$ has the form
    $$
    w^{(2)}_{e_1,j}(\lambda,x)=a^{(2)}_{e_1,j}(\lambda)\varphi_{e_1}(\lambda,x)+
    b^{(2)}_{e_1,j}(\lambda)S_{e_1}(\lambda,x),
    $$
where the constants $a^{(2)}_{e_1,j}(\lambda),b^{(2)}_{e_1,j}(\lambda)$ are unknown.
    
By the Kirchhoff conditions at $\gamma_2$, we have
\begin{equation}\label{0c2-1}
    \varphi_j(\lambda,\gamma_2)+M^{(2)}_{j,j}(\lambda)S_j(\lambda,\gamma_2)=c^{(2)}_{j,2}(\lambda),
\end{equation}
\begin{equation}\label{0cc2}
    a^{(2)}_{e_1,j}(\lambda)\varphi_{e_1}(\lambda,\gamma_2)+
    b^{(2)}_{e_1,j}(\lambda)S_{e_1}(\lambda,\gamma_2)=c^{(2)}_{j,2}(\lambda),
\end{equation}
\begin{equation}\label{0cc3}
    M^{(2)}_{j,k}(\lambda)S_k(\lambda,\gamma_2)=c^{(2)}_{j,2}(\lambda), \quad k\neq j,\ k=10,11,
\end{equation}
and
\begin{align}\label{0k-n2}
    &a^{(2)}_{e_1,j}(\lambda)\varphi'_{e_1}(\lambda,\gamma_2)+
b^{(2)}_{e_1,j}(\lambda)S'_{e_1}(\lambda,\gamma_2)+\varphi'_j(\lambda,\gamma_2)+\notag\\
    &\sum_{k=10,11}M^{(2)}_{j,k}(\lambda)S'_k(\lambda,\gamma_2)
    =c^{(2)}_{j,2}(\lambda)M^{(1)}_{2,2}(\lambda)+
    c^{(2)}_{j,7}(\lambda)M^{(1)}_{7,2}(\lambda),
\end{align}
where $M^{(1)}_{2,2}(\lambda)$ and $M^{(1)}_{7,2}(\lambda)$ are the elements of the Weyl matrix $M^{(1)}(\lambda): M^{(1)}_{2,2}(\lambda) = \partial w^{(1)}_2(\gamma_2)$, $M^{(1)}_{7,2}(\lambda)=\partial w^{(1)}_7(\gamma_2)$.

By the Kirchhoff conditions at $\gamma_7$, we have
\begin{equation}\label{r7c1}
    a^{(2)}_{e_1,j}(\lambda)\varphi_{e_1}(\lambda,\gamma_7)+
    b^{(2)}_{e_1,j}(\lambda)S_{e_1}(\lambda,\gamma_7)=c^{(2)}_{j,7}(\lambda),
\end{equation} 
\begin{equation}\label{r7c2}
    M^{(2)}_{j,k}(\lambda)S_k(\lambda,\gamma_7)=c^{(2)}_{j,7}(\lambda),\ k=8,9,
\end{equation}
and
\begin{align}\label{r7k}
    a^{(2)}_{e_1,j}(\lambda)\varphi'_{e_1}(\lambda,\gamma_7)+&b^{(2)}_{e_1,j}
    (\lambda)S'_{e_1}(\lambda,\gamma_7)+\sum_{k=8,9}M^{(2)}_{j,k}(\lambda)S'_k
    (\lambda,\gamma_7)\notag\\
    &=c^{(2)}_{j,2}(\lambda)M^{(1)}_{2,7}(\lambda)+c^{(2)}_{j,7}(\lambda)M^{(1)}
    _{7,7}(\lambda),
\end{align}
where $M^{(1)}_{2,7}(\lambda)=\partial w^{(1)}_2(\gamma_7)$, $M^{(1)}_{7,7}(\lambda)=\partial w^{(1)}_7(\gamma_7)$.

Thus, for each fixed $j \in \{ 10, 11\}$, from \eqref{0c2-1}-\eqref{r7k} we have eight equations with the eight unknowns $M^{(2)}_{j,k}(\lambda),k=8,\cdots,11$, $ c^{(2)}_{j,7}(\lambda)$, $c^{(2)}_{j,2}(\lambda)$, $a^{(2)}_{e_1,j}(\lambda)$, $ b^{(2)}_{e_1,j}(\lambda)$.
Note that in the linear algebraic system \eqref{0c2-1}-\eqref{r7k}, the magnitudes $\varphi_k (\lambda,x)$, $S_k(\lambda,x)$, $\varphi'_k(\lambda,x)$, $S'_k(\lambda,x)$ are known, since the potentials $q^{(2)}_k(x),k\in \{1,8,\cdots,11\}$ are known.
Then, we find $ M^{(2)}_{j,k}(\lambda),k=8,\cdots,11$, $a^{(2)}_{e_1,j}(\lambda)$, $b^{(2)}_{e_1,j}(\lambda)$, $c^{(2)}_{j,7}(\lambda)$ and $c^{(2)}_{j,2}(\lambda)$.

Note that here we obtain a larger linear system than in the synthesis of $M^{(1)}(\lambda)$ by $\tilde M(\lambda)$, because, by adding the edge $(\gamma_2, \gamma_7)$, we form a cycle. Therefore, we cannot add vertices to $\gamma_2$ and to $\gamma_7$ separately in order to decrease the size of linear systems. This simple example shows that the synthesis of the Weyl matrix for graphs with cycles differs from the case of tree-graphs.
	
Proceed with the construction of $M^{(2)}(\lambda)$. From \eqref{0w-w2}, we obtain 
\begin{equation*}
    M^{(2)}_{j,i}(\lambda)=c^{(2)}_{j,7}(\lambda)M^{(1)}_{7,i}(\lambda)+
    c^{(2)}_{j,2}(\lambda)M^{(1)}_{2,i}(\lambda),\ i=3,\cdots,6,\ j=10,11.
\end{equation*}

For $j=8,9$, we can similarly construct linear algebraic equations and obtain the elements $M^{(2)}_{j,i}(\lambda)$, $i=3,\cdots,6,8,\cdots,11$. Then, we have completed the rows $5,\cdots,8$ of the Weyl matrix $M^{(2)}(\lambda)$.

Now, choose an $i\in \{3,\cdots,6\}$. We look for $w^{(2)}_i(\lambda,x)$ such that on $\Gamma_0^{(1)}$ the following equality is valid:
$$
w^{(2)}_i(\lambda, x) = w^{(1)}_i(\lambda, x) + \alpha^{(2)}_{i,2}(\lambda)w^{(1)}_2(\lambda, x)+\alpha^{(2)}_{i,7}(\lambda)w^{(1)}_7(\lambda, x),
$$
where $\alpha^{(2)}_{i,j}(\lambda),j=2,7$ are constants. This is a natural choice, because
$$
w^{(2)}_i(\lambda,\gamma_i)=1,\  i=3,\cdots,6,
$$ 
and 
$$
w^{(2)}_i(\lambda,\gamma_j)=0,\  j=3,\cdots,6,\  j \neq i.
$$

By the Kirchhoff conditions at $\gamma_2$, we have
\begin{equation}\label{0cr2-2}
    M^{(2)}_{i,j}(\lambda)S_j(\lambda,\gamma_2)=\alpha^{(2)}_{i,2}(\lambda),\ j=10,11,
\end{equation}
\begin{equation}\label{0cr2-3}
    a^{(2)}_{e_1,i}(\lambda)\varphi_{e_1}(\lambda,\gamma_2)+
    b^{(2)}_{e_1,i}(\lambda)S_{e_1}(\lambda,\gamma_2)=\alpha^{(2)}_{i,2}(\lambda),
\end{equation}
and
\begin{align}\label{0kr2}
    &a^{(2)}_{e_1,i}(\lambda)\varphi'_{e_1}(\lambda,\gamma_2)+
    b^{(2)}_{e_1,i}(\lambda)S'_{e_1}(\lambda,\gamma_2)+\sum_{j=10,11}
    M^{(2)}_{i,j}(\lambda)S'_j(\lambda,\gamma_2)\notag\\
    &=M^{(1)}_{i,2}(\lambda)+\alpha^{(2)}_{i,2}(\lambda)M^{(1)}_{2,2}(\lambda)
    +\alpha^{(2)}_{i,7}(\lambda)M^{(1)}_{7,2}(\lambda).
\end{align}

By the Kirchhoff conditions at $\gamma_7$, we have
\begin{equation}\label{0cr7-2}
    M^{(2)}_{i,j}(\lambda)S_j(\lambda,\gamma_7)=\alpha^{(2)}_{i,7}(\lambda),\ j=8,9,
\end{equation}
\begin{equation}\label{0cr7-3}
    a^{(2)}_{e_1,i}(\lambda)\varphi_{e_1}(\lambda,\gamma_7)+
    b^{(2)}_{e_1,i}(\lambda)S_{e_1}(\lambda,\gamma_7)=\alpha^{(2)}_{i,7}(\lambda),
\end{equation}
and
\begin{align}\label{0kr7}
    &a^{(2)}_{e_1,i}(\lambda)\varphi'_{e_1}(\lambda,\gamma_7)+
    b^{(2)}_{e_1,i}(\lambda)S'_{e_1}(\lambda,\gamma_7)+
    \sum_{j=8,9}M^{(2)}_{i,j}(\lambda)S'_j(\lambda,\gamma_7)\notag\\
    &=M^{(1)}_{i,2}(\lambda)+\alpha^{(2)}_{i,2}(\lambda)M^{(1)}_{2,7}(\lambda)
    +\alpha^{(2)}_{i,7}(\lambda)M^{(1)}_{7,7}(\lambda).
\end{align}

For every $i\in \{3,\cdots,6\}$, the relations \eqref{0cr2-2}-\eqref{0kr7} give us eight  equations with the eight unknowns 
$$
\{M^{(2)}_{i,j}(\lambda),j=8,\cdots,11;
\alpha^{(2)}_{i,j}(\lambda),j=2,7,
a^{(2)}_{e_1,i}(\lambda),b^{(2)}_{e_1,i}(\lambda)\}.
$$
Then, we get the elements of the Weyl matrix $M^{(2)}(\lambda)$ in the columns $5,\cdots,8$ of the rows $1,\cdots,4$.

For $i, j = 3,\cdots,6$, we observe that
$$
\partial w^{(2)}_i(\lambda,\gamma_j)=\partial w^{(1)}_i(\lambda,\gamma_j)+ \alpha^{(2)}_{i,2}(\lambda)\partial w^{(1)}_2(\lambda,\gamma_j)+\alpha^{(2)}_{i,7}(\lambda)\partial w^{(1)}_7(\lambda,\gamma_j),
$$
and thus
\begin{equation}\label{0M_3}
  M^{(2)}_{i,j}(\lambda)=M^{(1)}_{i,j}(\lambda) + \alpha^{(2)}_{i,2}(\lambda)M^{(1)}_{2,j}(\lambda)+
  \alpha^{(2)}_{i,7}(\lambda)M^{(1)}_{7,j}(\lambda),\ i,j=3,\dots,6.
\end{equation}
Then, we construct the Weyl matrix $M^{(2)}(\lambda)=[M_{ij}^{(2)}(\lambda)]_{i,j=3,\ldots,6,8,\ldots,11}$ of the graph $\Gamma_1$ from the Weyl matrix $M^{(1)}(\lambda)$ of the graph $\Gamma_0^{(1)}$ by adding new edges to the right boundary vertices $\gamma_2$ and $\gamma_7$ of $\Gamma_0^{(1)}$.

Thus, we have obtained the Weyl matrix $M(\lambda)=M^{(2)}(\lambda)$ of the $1\ast 1$ square graph $\Gamma_1$. 

\section{General rectangular graph}\label{M-R}
\begin{figure}[ht]
		\centering
		\includegraphics[width=12cm,height=10cm]{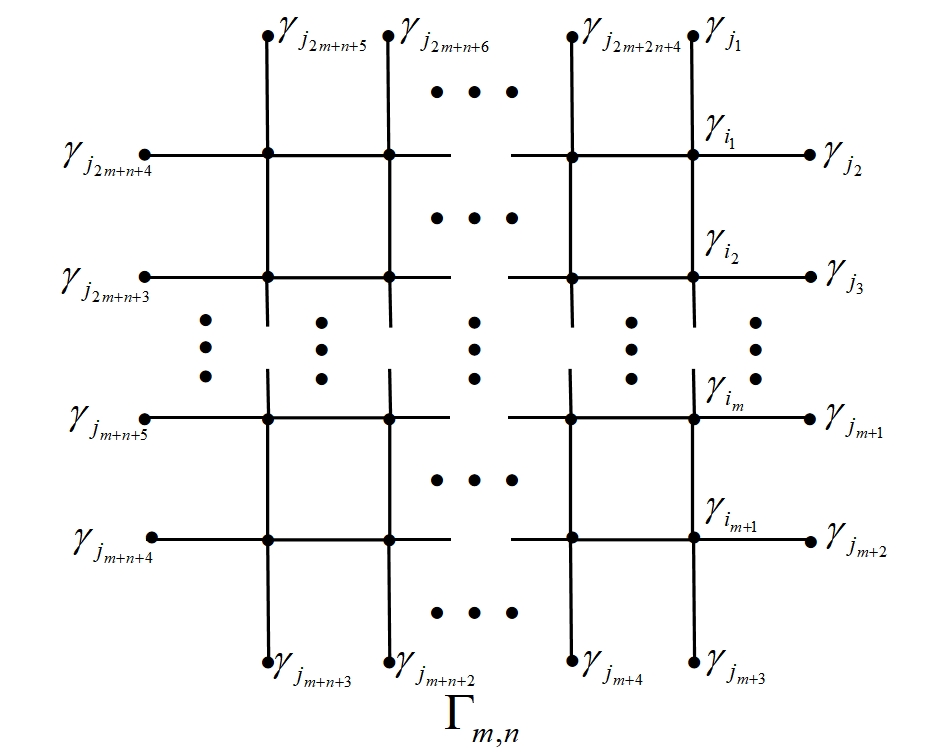}
		\caption{$m\ast n$ rectangular graph}
		\label{GN}
\end{figure}

\begin{figure}[ht]
		\centering
		\includegraphics[width=12cm,height=10cm]{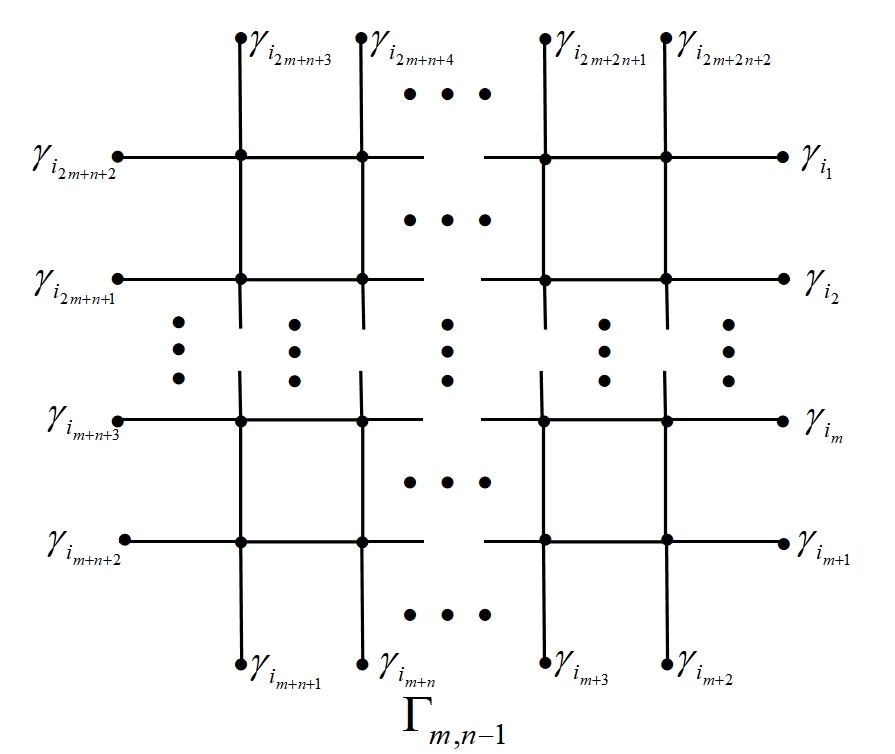}
		\caption{$m\ast (n-1)$ rectangular graph}
		\label{GN-1}
\end{figure}

In this section, our main purpose is to derive the procedure of the construction of the Weyl matrix of an arbitrary $m\ast n$ rectangular graph $\Gamma_{m,n}$ (see figure~\ref{GN}), $m,n \ge 0$. Clearly, $\Gamma_{m,0}$ for any $m$ is a tree, so the Weyl matrix for it can be obtained by the method of \cite{Avd4}. Therefore, we only need to develop a procedure for constructing the Weyl matrix $M(\lambda)=M^{(2)}(\lambda)$ from the Weyl matrix $M^{(1)}(\lambda)$ of $m\ast (n-1)$ rectangular graph $\Gamma_{m,n-1}$ (see figure~\ref{GN-1}). Then, this procedure can be applied inductively to pass from $\Gamma_{m,0}$ to $\Gamma_{m,n}$ with any $n \ge 1$.


Let us add new edges to the right boundary vertices of the graph $\Gamma_{m,n-1}$ and obtain the $m\ast n$ graph $\Gamma_{m,n}$. As in figures~\ref{GN-1} and~\ref{GN}, we denote the right boundary vertices of the graph $\Gamma_{m,n-1}$ by $\gamma_{i_1}$, $\gamma_{i_2}$, $\cdots$, $\gamma_{i_{m+1}}$ and the boundary vertices on the added edges by $\gamma_{j_1}$, $\gamma_{j_2}$, $\cdots$, $\gamma_{j_{m+3}}$. The rest boundary vertices of $\Gamma_{m,n}$ are denoted by $\gamma_{j_{m+4}}$, $\cdots$, $\gamma_{j_{2m+2n+4}}$. Each of the vertices $\gamma_{i_1}$ and $\gamma_{i_{m+1}}$ is incident to two added boundary edges and each of the vertices $\gamma_{i_2}$, $\cdots$, $\gamma_{i_m}$ is incident to only one added boundary edge. Then, the Weyl matrix $M(\lambda)=M^{(2)}(\lambda)$ can be constructed as follows. 

Fix $j\in\{j_1,\cdots,j_{m+3}\}$.
Let us look for $w^{(2)}_j(\lambda,x)$ in the form
\begin{equation}\label{'nw-w1}
    w^{(2)}_j(\lambda,x)=\sum_{k=1}^{m+1}c^{(2)}_{j,i_k}(\lambda)w^{(1)}_{i_k}(\lambda,x)\  on \ \Gamma_{m,n-1}.
\end{equation}
That is, on $\Gamma_{m,n-1}$ the Weyl solution $w^{(2)}_j(\lambda,x)$ is a linear combination of $w^{(1)}_{i_k}(\lambda, x)$, $k=1,\cdots,m+1$. In this case, it automatically satisfies the homogeneous Dirichlet condition at $\{\gamma_{j_{m+4}},\cdots,\gamma_{j_{2m+2n+4}}\}$, and we still need to fulfill the conditions $w^{(2)}_{j}(\lambda,\gamma_j)=1$ and $w^{(2)}_{j}(\lambda,\gamma_i)=0$ for $i \neq j$, $i\in\{j_1,\cdots,j_{m+3}\}$. The solutions on the added edges $e_{i_k}=(\gamma_{i_k},\gamma_{i_{k+1}})$, $k=1,\cdots,m$ have the form
$$
    w^{(2)}_{j,i_k}(\lambda,x)=a^{(2)}_{j,i_k}(\lambda)\varphi_{i_k}(\lambda,x)+
    b^{(2)}_{j,i_k}(\lambda)S_{i_k}(\lambda,x),
$$
where the contants $a^{(2)}_{j,i_k}(\lambda),b^{(2)}_{j,i_k}(\lambda)$, $k=1,\cdots,m$ are unknown.

Let the set of the subindices of the vertices connected to the vertex $\gamma_i$ be denoted by
$$
N_{\gamma_i}=\{k \colon \gamma_k\sim \gamma_i\}\  on\  \Gamma_{m,n},
$$
and
$$
\partial N_{\gamma_i}=\{k \colon \gamma_k \sim \gamma_i\  and\ \gamma_k \ is \ a \  boundary \  vertex  \}.
$$
    
For $j\in \partial N_{\gamma_{i_1}}$, by the Kirchhoff conditions at $\gamma_{i}$, $i\in\{i_1,i_{m+1}\}$, we have
\begin{equation}\label{'nc1}
    \varphi_j(\lambda,\gamma_i)+M^{(2)}_{j,j}(\lambda)S_j(\lambda,\gamma_i)
    =c^{(2)}_{j,i}(\lambda),
\end{equation}
\begin{equation}\label{'nc2}
    a^{(2)}_{j,t}(\lambda)\varphi_{t}(\lambda,\gamma_i)+
    b^{(2)}_{j,t}(\lambda)S_{t}(\lambda,\gamma_i)=c^{(2)}_{j,i}(\lambda),\ e_t \sim \gamma_i,\ t\in \{i_k\}_{k=1}^m,
\end{equation}
\begin{equation}\label{'nc3}
    M^{(2)}_{j,k}(\lambda)S_k(\lambda,\gamma_i)=c^{(2)}_{j,i}(\lambda),\  k\in \partial N_{\gamma_{i}},\  k\neq j,
\end{equation}
and
\begin{align}\label{'nk1}
    &a^{(2)}_{j,t}(\lambda)\varphi'_{t}(\lambda,\gamma_i)+
    b^{(2)}_{j,t}(\lambda)S'_{t}(\lambda,\gamma_i)+\varphi'_j(\lambda,\gamma_i)
    +\notag\\
    &\sum_{k\in \partial N_{\gamma_{i}}}M^{(2)}_{j,k}(\lambda)S'_k(\lambda,\gamma_i)=\sum_{k=1}^{m+1}c^{(2)}_{j,i_k}(\lambda)M^{(1)}_{i_k,i}(\lambda),\ e_t \sim \gamma_i,
\end{align}
where $M^{(1)}_{i_k,i}(\lambda) = \partial w^{(1)}_{i_k}(\gamma_i)$, $k=1,\cdots,m+1$.

By the Kirchhoff conditions at $\gamma_i$, $i\in\{i_2,\cdots,i_m\}$, we have 
\begin{equation}\label{'nc4}
    a^{(2)}_{j,t}(\lambda)\varphi_{t}(\lambda,\gamma_i)+
    b^{(2)}_{j,t}(\lambda)S_{t}(\lambda,\gamma_i)=c^{(2)}_{j,i}(\lambda),\ e_t \sim \gamma_i,
\end{equation} 
\begin{equation}\label{'nc5}
    M^{(2)}_{j,k}(\lambda)S_k(\lambda,\gamma_i)=c^{(2)}_{j,i}(\lambda),\ k\in \partial N_{\gamma_i},
\end{equation}
and
\begin{align}\label{'nk2}
    M^{(2)}_{j,k}(\lambda)S'_k(\lambda,\gamma_i)&+\sum_{e_t \sim \gamma_i}\left(a^{(2)}_{j,t}(\lambda)\varphi'_{t}(\lambda,\gamma_i)+
    b^{(2)}_{j,t}(\lambda)S'_{t}(\lambda,\gamma_i)\right)\notag\\
    &=\sum_{t=1}^{m+1}c^{(2)}_{j,i_t}(\lambda)M^{(1)}_{i_t,i}(\lambda),\ k\in \partial N_{\gamma_i}.
\end{align}

Thus, for each $j\in \partial N_{\gamma_{i_1}}$, from \eqref{'nc1}-\eqref{'nk2} we have $4m+4$ equations for the $4m+4$ unknowns $M^{(2)}_{j,k}(\lambda)$, $ k\in\{j_1,\cdots,j_{m+3}\}$, $c^{(2)}_{j,i_k}(\lambda)$, $k=1,\cdots,m+1$, $a^{(2)}_{j,i_k}(\lambda)$, $b^{(2)}_{j,i_k}(\lambda)$, $k=1,\cdots,m$. 

Note that in the linear algebraic system \eqref{'nc1}-\eqref{'nk2}, the magnitudes $\varphi_k (\lambda,x)$, $S_k(\lambda,x)$, $\varphi'_k(\lambda,x)$, $S'_k(\lambda,x)$ are known, since the potentials on the new edges $q^{(2)}_k(x)$, $k\in \{i_1,\cdots,i_m,j_1,\cdots,j_{m+3}\}$ are known.
Then, we can obtain $ M^{(2)}_{j,k}(\lambda)$, $k\in\{j_1,\cdots,j_{m+3}\}$, $a^{(2)}_{j,i_k}(\lambda)$, $b^{(2)}_{j,i_k}(\lambda)$ and $c^{(2)}_{j,i_k}(\lambda)$. 

From \eqref{'nw-w1}, we get
\begin{equation}\label{Mn2}
    M^{(2)}_{j,k}(\lambda)=\sum_{k=1}^{m+1}c^{(2)}_{j,i_k}(\lambda)M^{(1)}
    _{i_k,k}(\lambda),\ k\in\{j_{m+4},\cdots,j_{2m+2n+4}\}.
\end{equation}

For $j\in \partial N_{\gamma_{i}}$, $i\in\{i_2,\cdots,i_{m+1}\}$, we can similarly construct the linear algebraic equations and obtain the elements $M^{(1)}_{j,k}(\lambda)$, $k\in \{j_1,\cdots,j_{2m+2n+4}\}$. Thus, we have completed the rows $1,\cdots,m+3$ of the Weyl matrix $M^{(2)}(\lambda)$.

Now, choose an $i\in \{j_{m+4},\cdots,j_{2m+2n+4}\}$. We look for $w^{(2)}_i(\lambda,x)$ such that on $\Gamma_{m,n-1}$ the following equality is valid:
$$
w^{(2)}_i(\lambda, x) = w^{(1)}_i(\lambda, x) + \sum_{k=1}^{m+1}\alpha^{(2)}_{i,i_k}(\lambda)w^{(1)}_{i_k}(\lambda, x),
$$
where $\alpha^{(2)}_{i,i_k}(\lambda)$ are constants. This is a natural choice, because
$w^{(2)}_i(\lambda,\gamma_i) = 1$, $i\in \{j_{m+4},\cdots,j_{2m+2n+4}\}$ and $w^{(2)}_i(\lambda,\gamma_j) = 0$ for $j \in \{j_{m+4},\cdots,j_{2m+2n+4}\}$, $j \neq i$.

By the Kirchhoff conditions at $\gamma_k$, $k\in\{i_1,i_{m+1}\}$, we have
\begin{equation}\label{'4c1}
    M^{(2)}_{i,j}(\lambda)S_j(\lambda,\gamma_k)=\alpha^{(2)}_{i,k}(\lambda),\ j\in \partial N_{\gamma_k},
\end{equation}
\begin{equation}\label{'4c2}
    a^{(2)}_{i,t}(\lambda)\varphi_{t}(\lambda,\gamma_k)+
    b^{(2)}_{i,t}(\lambda)S_{t}(\lambda,\gamma_k)=\alpha^{(2)}_{i,k}(\lambda),\ e_t \sim \gamma_k,
\end{equation}
and
\begin{align}\label{'4k1}
    a^{(2)}_{i,t}(\lambda)&\varphi'_{t}(\lambda,\gamma_k)+
    b^{(2)}_{i,t}(\lambda)S'_{t}(\lambda,\gamma_k)+\sum_{j\in \partial N_{\gamma_k}}M^{(2)}_{i,j}(\lambda)S'_j(\lambda,\gamma_k)\notag\\
    &=M^{(1)}_{i,k}(\lambda) + \sum_{j=1}^{m+1} \alpha^{(2)}_{i,i_j}(\lambda)M^{(1)}_{i_j,k}(\lambda),\ e_t\sim\gamma_k.
\end{align}

By the Kirchhoff conditions at $\gamma_k$, $k\in\{i_2,\cdots,i_m\}$, we have
\begin{equation}\label{'4c1'}
    M^{(2)}_{i,j}(\lambda)S_j(\lambda,\gamma_k)=\alpha^{(2)}_{i,k}(\lambda),\ j\in \partial N_{\gamma_k},
\end{equation}
\begin{equation}\label{'4c2'}
    a^{(2)}_{i,t}(\lambda)\varphi_{t}(\lambda,\gamma_k)+
    b^{(2)}_{i,t}(\lambda)S_{t}(\lambda,\gamma_k)=\alpha^{(2)}_{i,k}(\lambda),\ 
    e_t \sim \gamma_k,
\end{equation}
and
\begin{align}\label{'4k1'}
    \sum_{e_t \sim \gamma_k}&\left(a^{(2)}_{i,t}(\lambda)\varphi'_{t}(\lambda,\gamma_k)+
    b^{(2)}_{i,t}(\lambda)S'_{t}(\lambda,\gamma_k)\right)+
    M^{(2)}_{i,j}(\lambda)S'_j(\lambda,\gamma_k)\notag\\
    &=M^{(1)}_{i,k}(\lambda) +\sum_{t=1}^{m+1} \alpha^{(2)}_{i,i_t}(\lambda)M^{(1)}_{i_t,k}(\lambda),\ j\in \partial N_{\gamma_k}.
\end{align}

For every $i\in \{j_{m+4},\cdots,j_{2m+2n+4}\}$, the relations \eqref{'4c1}-\eqref{'4k1'} give us $4m+4$ equations for the $4m+4$ unknowns 
$M^{(2)}_{i,j}(\lambda)$, $j\in\{j_1,\cdots,j_{m+3}\}$, $\alpha^{(2)}_{i,i_k}(\lambda)$, $k=1,\cdots,m+1$, 
$a^{(2)}_{i,i_k}(\lambda)$, $b^{(2)}_{i,i_k}(\lambda)$, $k=1,\cdots,m$.

So, we get the elements of the Weyl matrix $M^{(2)}(\lambda)$ in the columns $1,\cdots,m+3$ of the rows $m+4,\cdots,2m+2n+4$.

Finally, to obtain $M^{(2)}_{i,j}(\lambda)$ for $i, j \in\{ j_{m+4},\cdots,j_{2m+2n+4}\}$, that is, the elements of the Weyl matrix $M^{(2)}(\lambda)$ in the columns $m+4,\cdots,2m+2n+4$ of the rows $m+4,\cdots,2m+2n+4$. Observe that
$$
\partial w^{(2)}_i(\lambda,\gamma_j)=\partial w^{(1)}_i(\lambda,\gamma_j)+ \sum_{k=1}^{m+1}\alpha^{(2)}_{i,i_k}(\lambda)\partial w^{(1)}_{i_k}(\lambda,\gamma_j),
$$
and so
\begin{equation}\label{'4M_3}
  M^{(2)}_{i,j}(\lambda)=M^{(1)}_{i,j}(\lambda) + \sum_{k=1}^{m+1}\alpha^{(2)}_{i,i_k}(\lambda)M^{(1)}_{i_k,j}(\lambda)\ for \ i,j\in\{j_{m+4},\cdots,j_{2m+2n+4}\}.
\end{equation}
Thus, we have constructed the Weyl matrix $$M^{(2)}(\lambda)=[M_{ij}^{(2)}(\lambda)]_{i,j\in\{j_k\}_{k=1}^{2m+2n+4}}$$ of the graph $\Gamma_{m,n}$ from the Weyl matrix $M^{(1)}(\lambda)$ of the graph $\Gamma_{m,n-1}$ by adding new edges on the right boundary vertices.

Now, let us summarize the described procedure in the following algorithm.

\textbf{Algorithm 1.} 
Suppose that the Weyl matrix $M^{(1)}(\lambda) = [M_{ij}^{(1)}(\lambda)]_{i,j\in\{i_k\}_{k=1}^{2m+2n+2}}$ of the graph $\Gamma_{m,n-1}$ in figure~\ref{GN-1} is given. We have to construct the 
Weyl matrix $M^{(2)}(\lambda) = [M_{ij}^{(2)}(\lambda)]_{i,j\in\{j_k\}_{k=1}^{2m+2n+4}}$ of the graph
$\Gamma_{m,n}$ in figure~\ref{GN}. 

\smallskip

(1) For a fixed $j\in \partial N_{\gamma_{i_1}}$ in figure~\ref{GN}, solve the $(4m+4)\times(4m+4)$-system of linear algebraic equations \eqref{'nc1}-\eqref{'nk2} to find $ M^{(2)}_{j,k}(\lambda)$, $k\in\{j_1,\cdots,j_{m+3}\}$, and $c^{(2)}_{j,i_k}(\lambda)$, $k=1,\cdots,m+1$, $a^{(2)}_{j,i_k}(\lambda)$, $b^{(2)}_{j,i_k}(\lambda)$, $k=1,\cdots,m$. Then, we can compute $M^{(2)}_{j,k}(\lambda),k\in\{j_{m+4},\cdots,j_{2m+2n+4}\}$ by \eqref{Mn2}.

\smallskip

(2) For $j\in \partial N_{\gamma_{i}}$, $i\in\{i_2,\cdots,i_{m+1}\}$,  construct the linear algebraic equations similarly and obtain the entries $M^{(2)}_{j,k}(\lambda)$, $k\in \{j_1,\cdots,j_{2m+2n+4}\}$.

\smallskip

(3) For every $i\in \{j_{m+4},\cdots,j_{2m+2n+4}\}$, solve the $(4m+4)\times(4m+4)$-system 
of linear algebraic equations equations \eqref{'4c1}-\eqref{'4k1'} to find 
$M^{(2)}_{i,j}(\lambda)$, $j\in\{j_1,\cdots,j_{m+3}\}$, $\alpha^{(2)}_{i,i_k}(\lambda)$, $k=1,\cdots,m+1$, 
$a^{(2)}_{i,i_k}(\lambda)$, $b^{(2)}_{i,i_k}(\lambda)$, $k=1,\cdots,m$. Then, we can compute $M^{(2)}_{i,j}(\lambda)$, $j\in\{j_{m+4},\cdots,j_{2m+2n+4}\}$ by \eqref{'4M_3}.

\smallskip

Thus, all the elements of the Weyl matrix $M^{(2)}(\lambda)$ are constructed.

\medskip

Starting from the tree graph $G_{m,0}$ and applying Algorithm~1 inductively, we can obtain the Weyl matrix for any rectangular graph $G_{m,n}$, $m,n \ge 0$.
    
\section{Conclusions}\label{sec:con}
The procedure described above gives us a method for successive synthesis of a Weyl matrix of the $m\ast n$ rectangular subgraph of the square lattice. It allows one to compute the Weyl matrix by adding new edges and solving elementary systems of linear algebraic equations at each step. Due to the fact that the transposed Weyl matrix is the D-N map of the quantum graph, the synthesis procedure will find applications in solving a variety of boundary value and control problems on quantum graphs.

The correctness of the developed algorithm is based on the unique solvability of the linear algebraic systems at each step. Anyway, in practice, one needs to solve these systems only for particular values of $\lambda$. For example, in \cite{Avd3}, the authors use the Weyl matrix at a finite number of the points $\lambda$ to reconstruct the potentials on the quantum graph. The determinants of our linear systems are some non-zero meromorphic functions in $\lambda$, so a finite set of the spectral parameter values always can be chosen so that these systems are uniquely solvable. An example of analysis of the unique solvability for such systems is provided in Section~\ref{sec:Syn} (see Proposition~\ref{prop:solve}).

\section*{Conflicts of interests}
The author declared no potential conflicts of interest with respect to the research, author-ship, and publication of this article.
\section*{Data availability}
Data availability is not applicable to this article as no new data were created or analyzed in this study.

\end{document}